\documentclass[10pt,prd,superscriptaddress,preprintnumbers,amsmath,amssymb,nofootinbib]{revtex4}
\usepackage{epsfig}  
\usepackage{graphicx}
\usepackage{hyperref}
\usepackage{color}
\usepackage{float}
\usepackage{amsfonts}
\usepackage{amsmath}
\usepackage{slashed}

\large

\begin{document} 

\title{ Modeling the colors of phase noise in optomechanical oscillators}
\author{Cijy Mathai}
\affiliation{Indian Institute of Technology Bombay, Mumbai 400076, Maharashtra, India}

\author{Sunil A. Bhave}
\affiliation{OxideMEMS Laboratory, Purdue University, West Lafayette, IN, 47907 USA}

\author{Siddharth Tallur}
\email{stallur@ee.iitb.ac.in}
\affiliation{Indian Institute of Technology Bombay, Mumbai 400076, Maharashtra, India}

\begin{abstract}
Optomechanical oscillators (OMOs) combine the co-existing high quality factor mechanical and optical resonances in an integrated device to realize low phase noise RF oscillations. While several attempts have been demonstrated towards modeling the phase noise in such oscillators, the close-to-carrier phase noise models in literature do not account for $1/f^3$ (pink noise) and higher order slopes in the phase noise spectra. Here we present a phase noise model, corroborated with experimental characterization of phase noise of two monolithic integrated silicon 
OMOs, accounting for contributions to the phase noise due to thermomechanical, and adsorption-desorption (AD) noise. The model shows good agreement with experimental data and provides further insights into the mechanisms underlying the noise processes contributing to different slopes in the phase noise spectra in OMOs.
\end{abstract}
 
\maketitle 

OMOs exhibit self-sustained mechanical oscillations of cavity resonators 
possessing high quality factor mechanical and optical resonances \cite{Kippenberg:07}. 
The large spatial overlap of the mechanical and optical resonances generates optical back-action through 
the radiation pressure force, that can be tuned to be sufficiently large to overcome intrinsic mechanical
damping in the resonators, resulting in oscillations. Various quantum phenomena such as quadrature 
squeezing of light, parametric sideband and feedback cooling of mechanical oscillator close 
to its quantized zero point energy state, quantum phase fluctuations, etc. can be realized using 
OMOs \cite{Zhang:18, XuAntibunchingOMO2013, Eerkens:15}. 
The phase noise of OMOs is an important parameter that determines performance of these oscillators \cite{ST2010IFCS, Tallur:11} and should be accurately quantified for classical and quantum phase fluctuation experiments. The source of micro-mechanical oscillator phase noise can be attributed to thermal 
and mechanical fluctuations, photon shot noise, adsorption-desorption (AD) noise processes, and various other phenomena studied in the MEMS and NEMS communities \cite{Yang2011, Yong1989}. Phase noise modeling literature for OMOs \cite{ST2010IFCS, ManiPRA2006, HongTangPRA2014} does not adequately capture these noise sources, and consequently there are several demonstrations with higher order frequency slopes in the close-to-carrier phase noise spectra, that are not correctly accounted for \cite{ManiPRA2006,Luan2014}. In this paper, we measure and study the phase noise of two 
monolithic integrated silicon OMOs measured in a liquid nitrogen cooled vacuum chamber.
The phase noise spectrum exhibits different slopes, indicating that different sources of noise dominate in different 
offset frequency regimes. We derive an analytical model that captures the contribution of thermomechanical and AD noise sources. We observe
that AD noise processes dominate in vacuum at low temperatures, while thermomechanical noise sources dominate at ambient conditions.

\begin{figure}
\centering
    \includegraphics*[width=140mm]{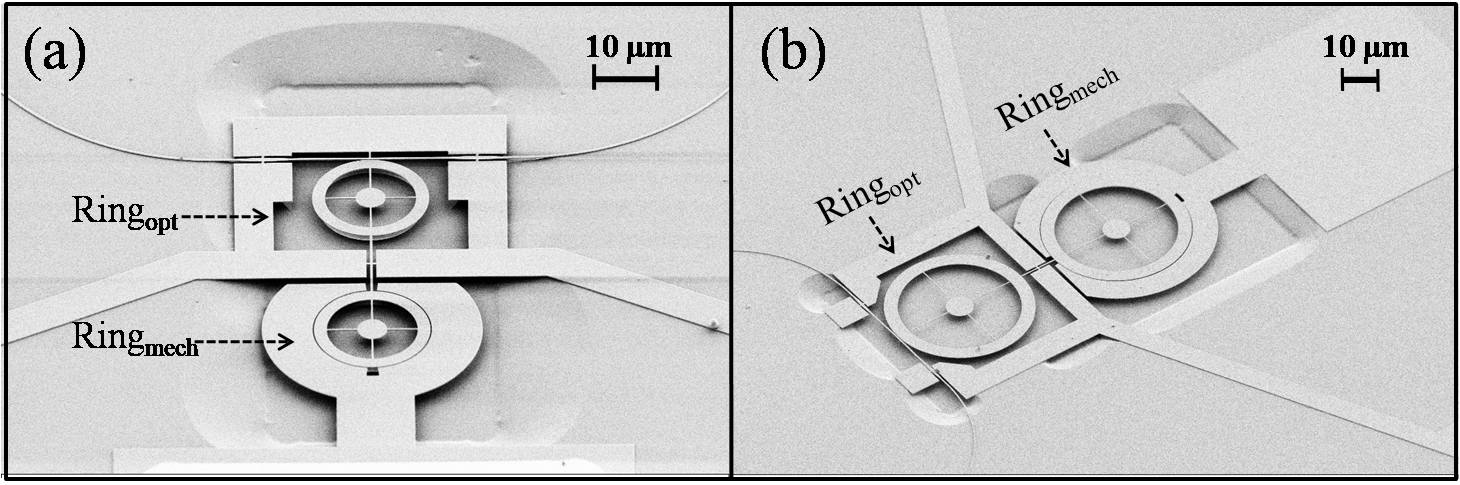}
    \caption{Scanning Electron Micrographs (SEM) of the two coupled ring silicon OMOs studied in this work. The rings differ in their lateral dimensions, as tabulated in Table \ref{parameter_table}. Panels (a) and (b) show the smaller (OMO$_1$) and larger (OMO$_2$) devices, respectively.}
\label{SEMs}
\end{figure}

\begin{figure}
\centering
    \includegraphics[width=100mm]{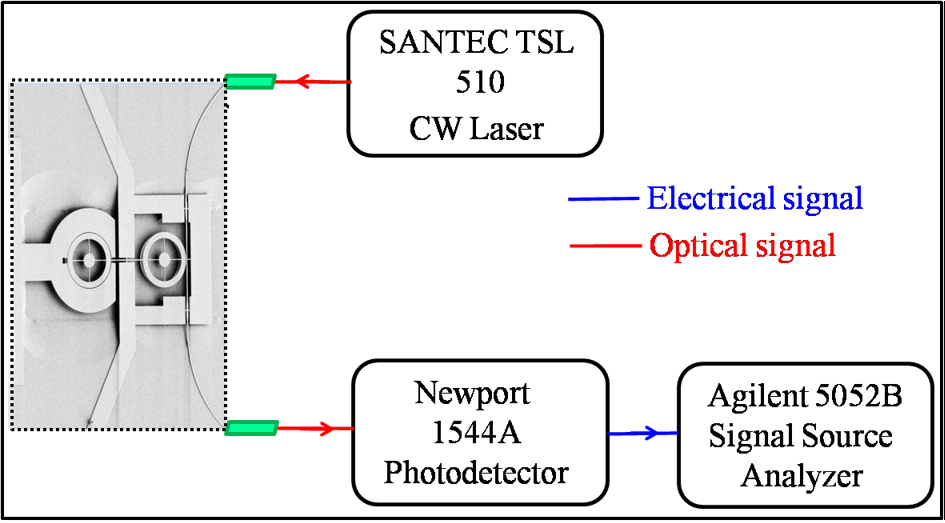}
    \caption{Block diagram of experimental setup used for characterizing the OMO phase noise.}
\label{setup1}
\end{figure}

Figure \ref{SEMs} shows the Scanning Electron Micrographs (SEM) of the optomechanical resonators (OMO$_1$ and OMO$_2$) used in our experiment, that are both coupled silicon ring resonators. The resonators are comprised of two coupled micro-mechanical rings that differ in their lateral dimensions (refer Table \ref{parameter_table}).
The fabrication process and design details of these ring resonators are presented in earlier work \cite{STIPJ2012}. Figure \ref{setup1} illustrates the experimental setup used to study the oscillator phase noise. The opto-mechanical ring resonator (labelled $Ring_{opt}$ in Figure \ref{SEMs} (a) and (b)) is optically coupled to the on-chip waveguide. Laser light from a tunable semiconductor diode laser operating in C- and L- bands (SANTEC TSL-510) is coupled to the waveguide using on-chip grating couplers and the laser wavelength is chosen such that it is blue-detuned to a high quality factor optical resonance of the resonator. 
Mechanical oscillations of the OMO excited by the radiation pressure force result in amplitude modulation of the light which is in turn converted into RF electrical signal using a Newport 1544-A photoreceiver connected at the output port of the waveguide. The oscillator phase noise is measured using an Agilent 5052B signal source analyzer. It is important to note that the two rings in an OMO are strongly coupled to each other through the mechanical coupling beam, and thus both rings undergo mechanical oscillations. The ring labeled $Ring_{mech}$ is used for capacitive transduction, when the device is operated with electrical feedback as an opto-acoustic oscillator \cite{STJMEMS2015}.

The oscillator frequency is affected by thermal fluctuations arising from additive white noise sources in the resonator.
The oscillation linewidth due to these thermal noise sources is represented as  
$\Delta \Omega = \frac{1}{2\pi}\bigg(\frac{4k_BT}{m_{eff}\Omega_0^2}\bigg)\bigg(\frac{\Delta \Omega_0}{r^2}\bigg),$
where $k_B$ is the Boltzmann constant, $T$ is the temperature, $m_{eff}$ is the effective mass of the resonator,
$\Omega_0$ (= 2$\pi f_0$) is the angular resonant frequency of the mechanical oscillator, 
$\Delta \Omega_0$ (= $\Omega_0/Q_{mech}$) is the intrinsic mechanical oscillation linewidth, $Q_{mech}$ is the quality factor of the mechanical resonance, and $r$ is the 
radial displacement amplitude of the optomechanical oscillator corresponding to the radiation pressure force \cite{ST2010IFCS, ManiPRA2006}. 
Therefore, the expression for thermomechanical phase noise density function can be represented using the Leeson model 
for electronic oscillators \cite{Leeson1966} as follows:
\begin{equation}
\label{SphiTN}
S_{\phi TN}(f) =  \\    
\begin{cases}
  10 \log_{10} \bigg[\frac{A^2}{2\pi}\frac{\Delta \Omega}{\Delta \Omega^2+(2\pi f)^2}\bigg], & f < \frac{\Delta \Omega_0}{2\pi} \\    

  10 \log_{10} \bigg[\frac{A^2}{2\pi}\frac{\Delta \Omega}{\Delta \Omega^2+\Delta \Omega_0^2}\bigg], & f \geq \frac{\Delta \Omega_0}{2\pi}    
\end{cases}
\end{equation}
where $A^2/2$ is the signal power at oscillation frequency.

The impact of adsorption, desorption, and diffusion of ambient gaseous flux on phase noise in micromechanical resonators has been studied systematically, through pioneering work by Yong and Vig \cite{Yong1989}. The ambient gas molecules adsorbed on resonator surfaces are highly localized and can interact with each other, facilitating multilayer adsorption. Extremely localized adsorption occurs due to the dominance of the strong surface potential over the kinetic energy of the adsorbed molecule. In our model we assume that each site on the resonator surface is occupied by only one gas molecule. Following Langmuir adsorption model \cite{Langmuir1918}, considering Maxwell-Boltzmann distribution and partition function of finite number of adsorbates in a canonical ensemble, we have the adsorption rate of each site represented as $r_a = \frac{P}{\sqrt{2\pi Mk_BT}}sA_{site}$, where $P$ and $T$ are the ambient pressure and temperature, $M$ is the mass of the adsorbed molecule, $s$ is the sticking coefficient of the adsorbed molecule ($0\leq s\leq1$), and $A_{site}$ is the area of the adsorbed site. In our experiment, the resonator is surrounded by air, and hence we assume $A_{site}= \pi r_{N_2}^2$, where $r_{N_2} = 0.188 nm$ is the Lennard-Jones radius of $N_2$ molecule \cite{MarcusLennardJonesRadius2003}. The molecular desorption rate from the surface can be expressed in terms of Arrhenius equation $r_d = \nu_d e^{-E_a/k_BT}$, where $k_B$ is the Botzmann constant, $\nu_d$ is the desorption attempt frequency \cite{Yong1989}, and $E_a$ is the binding energy of the adsorbed molecule.

Adsorption-desorption (AD) can be modeled as a random stochastic process occurring on the resonator surface, represented as the summation of time dependent Bernoulli random variable function $b_i(t)$ corresponding to each adsorption site $i$ \cite{BAlan1986}. 
The number of adsorption sites can be represented as $N = aN_t$, where $a$ and $N_t$ are the adsorption probability and the total number of sites on the oscillator surface, respectively.
If a site $i$ is occupied, the corresponding Bernoulli random variable $b_i(t) = 1$, and if unoccupied $b_i(t) = 0$. These random events are assumed to be stationary and mutually independent.  
\begin{figure}[h]
\centering
    \includegraphics*[width=140mm]{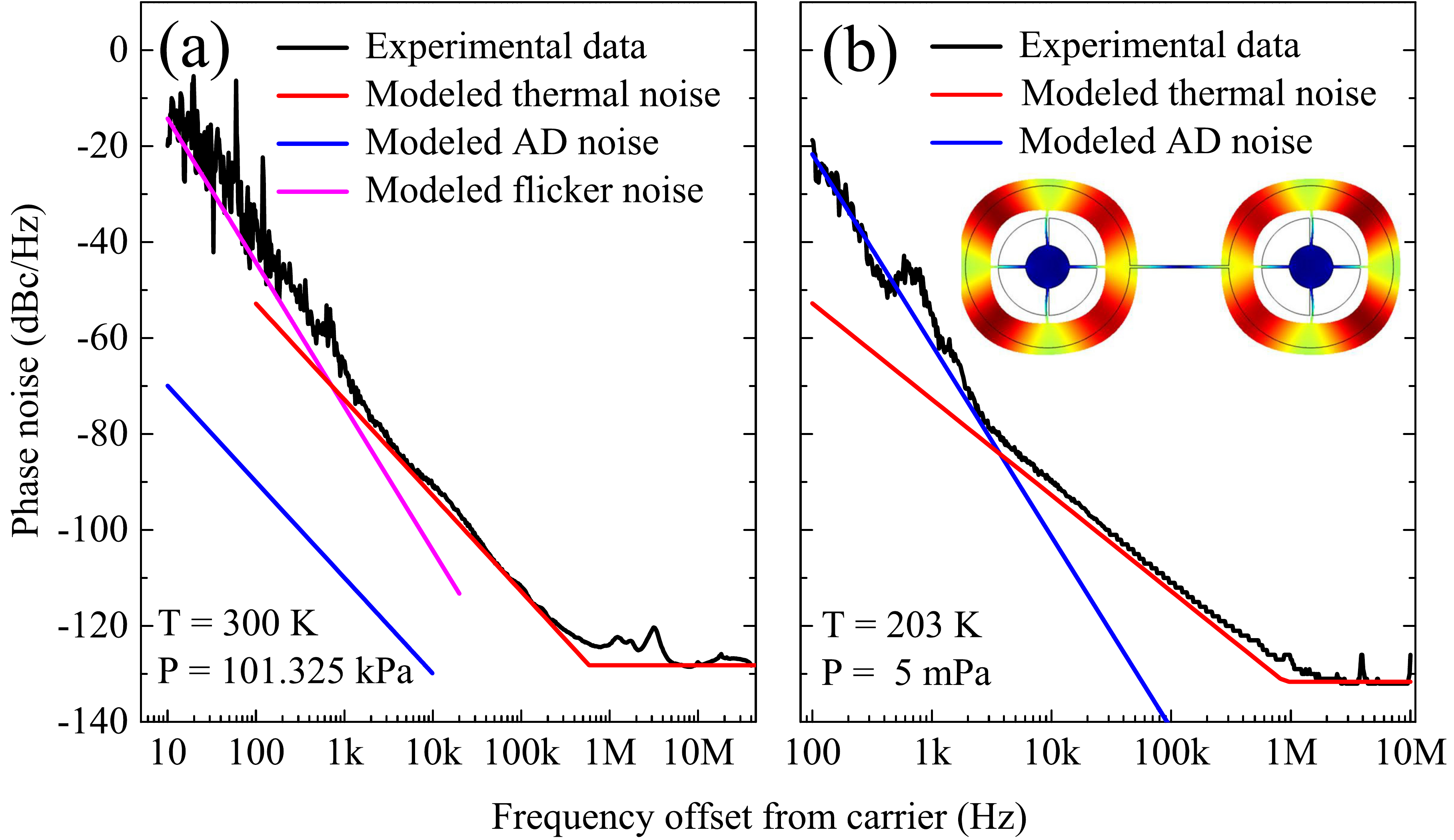}
    \caption{Phase noise spectra for the radial mode for OMO$_1$ (oscillation frequency 176 MHz) at (a) room temperature and pressure (with external electrical feedback using DC bias $V_{dc} = +17 V$; RF signal power, $P_{RF} = -18 dBm$; laser wavelength, $\lambda_0 = 1595.30 nm$), and (b) at low temperature in vacuum ($P_{RF} = -13.7 dBm$; $\lambda_0 = 1595.30 nm$). The inset in panel (b) shows the mode shape of the radial breathing mode simulated in COMSOL Multiphysics FEM software.}
\label{PinkBrown}
\end{figure}
These random processes cause mass fluctuation, and in turn lead to frequency fluctuations of the micromechanical resonator. The net frequency fluctuation caused by AD process is given by $f(t) = \sum_{i=1}^{N} \xi_ib_i(t)$, where $\xi_i$ (=$\frac{\Delta f}{N}$, and $\Delta f = f_0\big(\frac{\rho_A}{\rho_{Aeff}}\big)$). Here $\rho_A$ is the mass per unit area of the 
adsorbed molecule, and $\rho_{Aeff}$ is the effective mass per unit area of the resonator.\cite{Yong1989}. 
In order to obtain an expression for spectral phase noise density function, 
we need to consider the autocorrelation function $R(t)= E[f(t+\tau)f(t)]$, 
where $E[x]$ represents the expectation value of the function $x$, and $\tau$ represents the time lapse between two AD processes.  A random physical process such as the Bernoulli random process taking place between time
interval $t$ and $(t+\tau)$ can be represented in terms of Poisson distribution function, 
whose probability mass function is $P(A_n) = \frac{(\lambda|\tau)^n}{n!}e^{-\lambda|t|}$. 
Therefore, the autocorrelation function $R(t)$ can be represented as:
\begin{equation}
    R(t) =  \sum_{i=1}^{N}\xi_i^2E[ b_i(t+\tau)b_i(t)] = \sigma^2\sum_{n=0}^{\infty}(-1)^nP(A_n)
            = \sigma^2e^{-2\lambda|t|}.
\end{equation}
Here $2\lambda = \frac{1}{2\tau_r}$, where $\tau_r$ ($= \frac{1}{r_a+r_d}$) is the correlation time, 
and $\sigma^2$ represents the variance of the frequency fluctuation function $f(t)$, and can be written as:
\begin{equation}
    \sigma^2 = p(1-p)\sum_{i=1}^{N}\xi_i^2 = p(1-p)\frac{(\Delta f)^2}{N}
\end{equation}
where $p = \frac{r_a}{r_a+r_d}$ is the probability of a site to be occupied when adsorption occurs.
The spectral density function $S_f(f)$ corresponding to frequency noise can be obtained by taking the Fourier transform of the autocorrelation function $R(t)$ following Weiner-Khinchin theorem \cite{wiener1930},  which yields 
\begin{equation}
    S_f(f) =  F[\sigma^2e^{-|t|/\tau_r}] = \frac{2\sigma^2\tau_r}{1+(2\pi f\tau_r)^2}
\end{equation}

The spectral phase noise density function is expressed in terms of the frequency noise as $S_{\phi ADN}(f) = \frac{1}{f^2}S_f(f)$, and can be represented as
\begin{equation}\label{SphiAD}
S_{\phi ADN}(f) = \frac{2r_ar_d/N}{(r_a+r_d)^3+(2\pi f)^2(r_a+r_d)}\frac{f_0^2}{f^2}\bigg(\frac{\rho_A}{\rho_{Aeff}}\bigg)^2.
\end{equation} 
\begin{figure}[h]
\centering
    \includegraphics*[width=140mm]{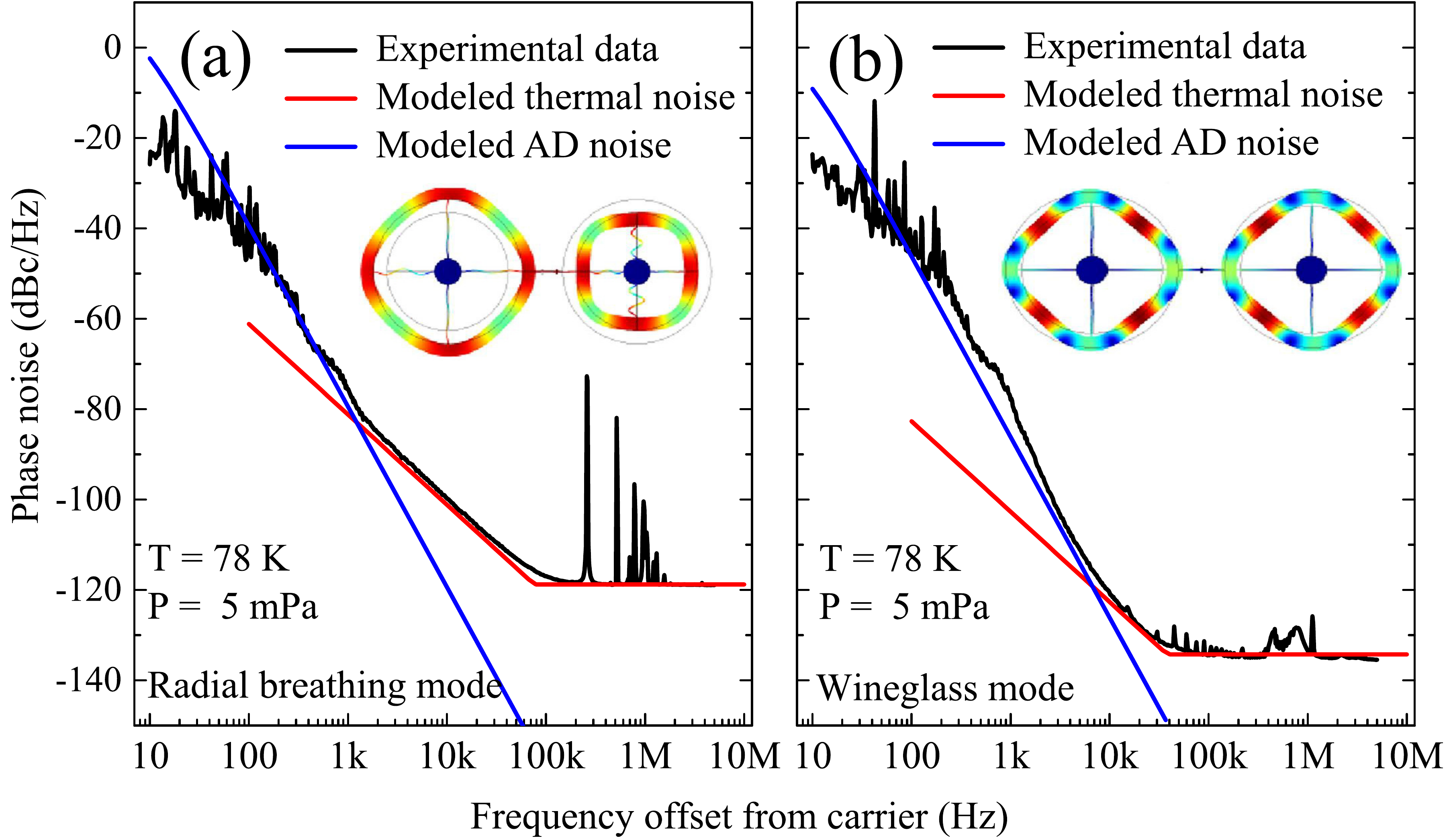}
    \caption{Phase noise spectra of OMO$_2$ for (a) radial breathing mode at oscillation frequency $70 MHz$ ($P_{RF} = -24.62 dBm$; $\lambda_0 = 1590 nm$), and (b) wineglass mode at frequency $58 MHz$ ($P_{RF} = -9 dBm$; $\lambda_0 = 1590 nm$). Insets show the corresponding mode shapes simulated in COMSOL Multiphysics FEM software.}
\label{WG_rad}
\end{figure}
We have obtained the phase noise spectra for two coupled ring resonator systems, OMO$_1$ and OMO$_2$ (see Fig. \ref{SEMs}). 
OMO$_1$ has been used to study the environmental influence (such as change in ambient temperature and pressure) on oscillators, 
whereas OMO$_2$ has been studied to understand how effectively different mechanical modes (like wineglass mode and radial breathing mode) 
affect an oscillator. The resonators have high quality factor optical resonances exceeding $30,000$ in the C-band in vicinity of $1550nm$. The TSL-510 laser wavelength is blue detuned to a high quality factor optical resonance to excite radiation pressure driven self-sustained mechanical oscillations in these resonators.
The radial breathing mode with frequency 176MHz \cite{STIPJ2012} is studied for OMO$_1$ at room pressure and temperature and also at low pressure and temperature, while for OMO$_2$ two different modes are studied at low pressure and temperature, namely the radial breathing mode and wineglass mode \cite{STJMEMS2015}. The goal of the OMO$_1$ study is to understand which noise source (thermomechanical vs. AD) is dominant at these vastly different operating ambient conditions, and the goal of the OMO$_2$ study is to understand how different mechanical modes are affected by AD phase noise. The measurement conducted on OMO$_1$ for room temperature ($300K$) and pressure ($101.325 kPa$) required a modification to the experimental setup shown in Figure \ref{setup1}. An external feedback loop \cite{STJMEMS2015} is required to launch mechanical oscillations due to the high threshold power for radiation pressure driven oscillations. 
The corresponding phase noise spectrum is shown in Figure \ref{PinkBrown}(a). The amplifier flicker (pink) noise and the thermomechanical noise models agree well with the experimental data 
(see Table. \ref{parameter_table} for all parameter values used for the phase noise model). For the theoretical fitting, we have considered $1 kcal/mol\leq E_a \leq 12 kcal/mol$ \cite{Yang2011, Kritzenberger1994, Rittner1995} , $10 kHz \leq \nu_d \leq 10 THz$ \cite{Yang2011}, and $0\leq a \leq 1$ as constraints for the curve fits.
The cubic term in the denominator term of equation (\ref{SphiAD}) dominates at room temperature and pressure, and hence the contribution of the AD phase noise appears as $1/f^2$ and has $4$ orders of magnitude lower contribution to the phase noise as compared to the amplifier flicker noise and thermomechanical noise. At low temperature ($203K$) and pressure ($5mPa$), the phase noise spectrum shows $1/f^4$ dependence on the frequency offset from carrier, which fits well with the modeled AD phase noise as shown in Figure \ref{PinkBrown}(b). The external feedback loop is not necessary for low temperature, as the mechanical quality factor is sufficiently high to lower the threshold optical power required to launch self sustained oscillators, within the operating range of the TSL-510. 
\begin{table*}[ht!]
\centering
\caption{\label{parameter_table} 
Parameters used for the phase noise models. The sticking coefficient ($s$) is assumed to be equal to $1$ for all cases.}
\begin{tabular}{|p{3cm}|p{2cm}|p{2cm}|p{2cm}|p{2cm}|p{4.5cm}|}
\hline
 \textbf{Parameters} & \multicolumn{2}{c|}{\textbf{OMO$_1$ (Radial mode)}} & \textbf{OMO$_2$ (Radial mode)}
 &\textbf{OMO$_2$ (Wineglass mode)}&\textbf{ Remarks}\\ \hline
  Dimension of the ring resonator & \multicolumn{2}{c|}{$rad_{out} = 10 \mu m$, $rad_{in} = 6.2 \mu m$}&
   \multicolumn{2}{c|}{$rad_{out} = 21 \mu m$, $rad_{in} = 17.2 \mu m$} & As designed, and confirmed with SEM images \\ \hline
  Effective mass of the resonator (m$_{eff}$) & \multicolumn{2}{c|}{$116.56 pg$} &$264.35 pg$ &$176.23 pg$  & Extracted from mechanical eigenmode simulations using COMSOL Multiphysics\\ \hline
   Pressure ($P$)&$5 mPa$ &$101.325 kPa$ & \multicolumn{2}{c|}{$5 mPa$} & Ambient pressure\\ \hline 
   Temperature ($T$)&$203 K$ & $300 K$ &\multicolumn{2}{c|}{$78 K$} & Ambient temperature\\ \hline
    Mechanical resonance frequency ($f_0$) &\multicolumn{2}{c|}{$176 MHz$} &$70 MHz$ & $58 MHz$ & Measured electro-mechanical transmission spectrum \cite{STIPJ2012, ST_Transducers2015}\\ \hline
  
    Mechanical resonance quality factor ($Q_{mech}$) &100 & 150 & 460 & 760 & Measured electro-mechanical transmission spectrum \\ \hline
    Desorption attempt frequency ($\nu_d$)& $8.69 THz$ & $9.10 THz$ &\multicolumn{2}{c|}{$5.12 THz$}  & Fitting parameter \\ \hline
    Binding energy ($E_a$)& \multicolumn{2}{c|}{$10 kcal/mol$}& \multicolumn{2}{c|}{$4 kcal/mol$}& Fitting parameter \\  \hline
    Adsorption probability (a)& \multicolumn{2}{c|}{$0.001$} & $0.05$ & $0.01$ & Fitting parameter\\ \hline
\end{tabular}
\end{table*}

The adsorbates are considered to be 
nitrogen molecules in the model, as nitrogen accounts for $78\%$ of the air inside the chamber. 
The surface area of the device (consisting of top and bottom surfaces of two rings) exposed to the impinging molecules 
is $\approx 4 \pi(r_{out}^2-r_{in}^2)$. This area can accommodate a tightly packed monolayer 
of $N_t \approx 7.3 \times 10^9$ $N_2$ molecules (wherein every molecule occupies an area $=\pi r_{N2}^2$; $r_{N2} = 0.188 nm$ is the Lennard-Jones radius of the nitrogen molecule \cite{MarcusLennardJonesRadius2003}). Here, $N (= aN_t)$ can be considered as the number of available adsorption sites, where $a$ is the adsorption probablity and 
$N_t$ is the total number of sites in the resonator. 
The rate of adsorption is $r_a = sI$, where $I$ ($= \frac{P}{\sqrt{2\pi Mk_BT}}A_{site}$) is the impinging rate of the $N_2$ molecules 
adsorbed on the device surface per unit area per unit time. The sticking coefficient $s$ is assumed to be equal to $1$,
as conventionally assumed for impinging of inert atoms/molecules on cold surfaces \cite{Raeker1998}. 
For the desorption rate of the atoms leaving the device surface, we follow the temperature dependent Arrhenius equation where the activation energy corresponding to surface diffusion process is far less than that of desorption process.
The modeled AD phase noise model fits the experimental data well for $E_a = 10 kcal/mol = 434 meV$ and $\nu_d \approx 9 THz$. 
The best-fit $E_a$ value is significantly greater than the surface diffusion energy reported for many inert 
gas-dielectric interaction surfaces with $E_{diff}< 100 meV$ \cite{Thomas2003, Yang2011}. Moreover, the best fit $E_a$ value also lies in the range of $200 meV-500 meV$, thereby agreeing with the desorption 
activation energy values reported in many systems involving inert gas-dielectric AD interactions \cite{Yang2011, Kritzenberger1994, Rittner1995}. This suggests that the dominant mechanism at play in these experiments is 
desorption rather than the surface diffusion.

To study the AD phase noise dependence on different mechanical modes (namely, wineglass and radial modes), we have 
performed the experiment with OMO$_2$ that has lower mechanical resonance frequencies and thereby lower threshold optical power for radiation pressure driven self-sustained oscillations \cite{ManiPRA2006}. 
The phase noise spectra for both modes measured at $T=78K$ have been shown in Figure \ref{WG_rad}. 
The modeled thermomechanical (equation \ref{SphiTN}) and AD (equation \ref{SphiTN}) phase noise agree well with the experimental data for parameters provided in Table \ref{parameter_table}. 
The effective mass of the wineglass mode is roughly one-third the mass of the coupled-rings, whereas 
the effective mass of the radial breathing mode is approximately half of the total mass of the coupled-ring resonator \cite{AMBATI1976415}. 
The best-fit adsorption probability $a$ differs for both modes, and the best-fit adsorption binding energy $E_a$ is $4 kcal/mol = 170 meV$ for both modes, which suggests that desorption dominates compared to lateral diffusion of adsorbed molecules \cite{Yang2011}.

In conclusion, we present a model for contribution to optomechanical oscillator phase noise due to adsorption-desorption of ambient gas molecules on the resonator surface, following phase noise models reported for MEMS and NEMS oscillators \cite{Yong1989,Yang2011}. The model is validated with experimentally measured phase noise spectra for two coupled silicon micro-mechanical ring oscillators, for two different mechanical modes (wineglass and radial breathing modes). The phase noise model presented in this work supplements previously reported models for opto-mechanical oscillators \cite{ST2010IFCS, ManiPRA2006, HongTangPRA2014} and explains previously unexplained source of $1/f^4$ phase noise in optomechanical oscillators \cite{ManiPRA2006,Luan2014}.

\textbf{ACKNOWLEDGMENT:}
The micro-mechanical devices reported in this work were fabricated at the Cornell NanoScale Science $\&$ Technology Facility (CNF). The authors thank Prof. Swati Singh at the University of Delaware for insightful discussions on origin of phase noise in optomechanical oscillators.

\end{document}